\documentclass[conference]{IEEEtran}
\IEEEoverridecommandlockouts
\usepackage{cite}
\usepackage{amsmath,amssymb,amsfonts}
\usepackage{algorithmic}
\usepackage{graphicx}
\usepackage{textcomp}
\usepackage{xcolor}
\usepackage{array}
\usepackage{stfloats}

\def\BibTeX{{\rm B\kern-.05em{\sc i\kern-.025em b}\kern-.08em
    T\kern-.1667em\lower.7ex\hbox{E}\kern-.125emX}}
\begin{document}

\title{Speech Augmentation Based Unsupervised Learning for Keyword Spotting}
\author{
\IEEEauthorblockN{Jian Luo$^1$, Jianzong Wang$^{1*}$\thanks{*Corresponding author: Jianzong Wang, jzwang@188.com}, Ning Cheng$^1$, Haobin Tang$^{1,2}$, Jing Xiao$^1$}
\IEEEauthorblockA{$^1$\textit{Ping An Technology (Shenzhen) Co., Ltd.}\\$^2$\textit{University of Science and Technology of China}}
}

\maketitle

\begin{abstract}
In this paper, we investigated a speech augmentation based unsupervised learning approach for keyword spotting (KWS) task. KWS is a useful speech application, yet also heavily depends on the labeled data. We designed a CNN-Attention architecture to conduct the KWS task. CNN layers focus on the local acoustic features, and attention layers model the long-time dependency. To improve the robustness of KWS model, we also proposed an unsupervised learning method. The unsupervised loss is based on the similarity between the original and augmented speech features, as well as the audio reconstructing information. Two speech augmentation methods are explored in the unsupervised learning: speed and intensity. The experiments on Google Speech Commands V2 Dataset demonstrated that our CNN-Attention model has competitive results. Moreover, the augmentation based unsupervised learning could further improve the classification accuracy of KWS task. In our experiments, with augmentation based unsupervised learning, our KWS model achieves better performance than other unsupervised methods, such as CPC, APC, and MPC.
\end{abstract}


\section{Introduction}
Keyword Spotting (KWS) is a useful speech application in real-world scenarios. KWS aims at detecting a relatively small set of pre-defined keywords in an audio stream, which usually exists on the interactive agents. The KWS systems usually have two kinds of applications: Firstly, it can detect the startup commands, such as ``hey Siri'' or ``OK, Google'', providing explicit cues for interactions. Secondly, KWS can help to detect some sensitive words to protect the privacy of the speaker. Therefore, highly accurate and robust KWS systems can be of great significance to real speech applications \cite{li2017acoustic, Luo2021End, schalkwyk2010your}.

Recently, extensive literature research on KWS has been published~\cite{wang2019adversarial,rosenberg2017end,shan2018attention}. As a traditional solution, keyword/filler Hidden Markov Model (HMM) has been widely applied to KWS tasks, and remains competitive results~\cite{silaghi2005spotting}. In this generative approach, an HMM model is trained for each keyword, while another HMM model is trained from not-keyword speech segments. At inference, the Viterbi decoding is required, which might be computationally expensive depending on the HMM topology. In recent years, deep learning models have gained popularity on the KWS task, which show better performance than traditional approaches. Google proposed to use Deep Neural Networks (DNN) to predict sub-keyword targets. It uses the posterior processing method to generate the final confidence score, and outperforms the HMM-based system~\cite{chen2014small}. In contrast, Convolutional Neural Networks (CNN) is more attractive, because DNN ignores the input topology, but audio features could have a strong dependency in time or frequency domains \cite{tang2018deep, Luo2021Unidirectional, Xu2020}. However, there is a potential drawback that CNN might not model much contextual information. Also, Recurrent Neural Networks (RNN) with Connectionist Temporal Classification (CTC) loss was also investigated for KWS. However, the limitation of RNN is that it directly models the speech features without learning local structure between successive time series and frequency steps~\cite{sun2016max}. There are also some works that combined CNN and RNN to improve the accuracy of KWS. For example, Convolutional Recurrent Neural Networks (CRNN) and Gated Convolutional Long Short-Term Memory (LSTM), achieved better performance than that of only using CNN or RNN~\cite{arik2017convolutional}. In recent years, many researchers focus on the transformer-based models with self-attention mechanism. As a typical model, Bidirectional Encoder Representations from Transformer (BERT) has been proven to be an effective model in many Natural Language Processing (NLP) tasks~\cite{vaswani2017attention, Jia2020Large, devlin2018bert}. The transformer-based models have also obtained much application in Automatic Speech Recognition (ASR) tasks~\cite{karita2019comparative, Luo2021Multi}. In this work, we introduced transformer to the network architecture of KWS. We think that transformer encoder has great advantage on the speech representation, and established a CNN-Attention based network to deal with the KWS task. The CNN helps network to learn the local feature, and the self-attention mechanism of transformer focuses on the long-time information.

The above supervised approaches have acquired good performance, but these models require a lot of labeled datasets. Obviously, for KWS task, the negative samples could be more procurable than positive samples, meaning that the positive samples might not be obtained easily. Especially when the keyword changes, it requires much time to collect the positive target samples, and the existing models might not easily transfer to other KWS models. In this paper, we focus on the unsupervised learning approach to alleviate this problem. The unsupervised learning mechanism allows the neural network to be trained on unlabeled datasets. With unsupervised learning, the performance of downstream task could be improved with limited labeled datasets. Unsupervised learning has made great success in the audio, image and text tasks~\cite{gi}. In speech area, researchers also proposed some unsupervised pre-training algorithms~\cite{wang2020self,Luo2021Dropout,schneider2019wav2vec}. Contrastive Predictive Coding (CPC) is one of those unsupervised approaches, extracting speech representation by predicting future information~\cite{oord2018representation}. Apart from CPC, the Autoregressive Predictive Coding (APC) is another pre-training model, which also gets comparable results on phoneme classification and speaker verification tasks~\cite{chung2019unsupervised}. Meanwhile, Masked Predictive Coding (MPC) designs a Masked Language Model (MLM) objective in the unsupervised pre-training, and enables the model to incorporate context from both directions~\cite{Improving}. Based on these unsupervised learning methods, lots of unlabeled audio data can be used to obtain a better audio representation and this representation can be applied to the follow-up tasks through fine-tuning mechanism. For a robust KWS system, it should deal with different styles of speech in real-world applications. Speed and volume are major variations of the speech style. Unlike traditional unsupervised learning focuses on the general audio representation, we proposed an augmentation based approach. Our approach is to improve the model performance on KWS task with different speed and intensity situations. We designed an unsupervised loss based on the distance between the original and augmented speech, as well as the audio reconstructing information for auxiliary training. We think that speech utterances with the same keyword but at different speeds or volumes should have similar high-level feature representations for KWS tasks.

This paper investigated unsupervised speech representative methods to conduct KWS task. The unsupervised learning methods could utilize a lot of unlabeled audio datasets to improve the performance of downstream KWS task when labeled data are limited. In addition, speech augmentation based unsupervised representation might help the network to learn the speech information in various speech styles, and get a more robust performance.
In summary, our major contributions of this work are the followings:
\begin{itemize}
	\setlength{\itemsep}{0pt}
	\setlength{\parsep}{0pt}
	\setlength{\parskip}{0pt}
	\item Propose a CNN-Attention architecture for keyword spotting task, having competitive results on Google Speech Commands V2 Dataset.
	\item Design an unsupervised loss based on the Mean Square Error (MSE) to measure the distance between the original and augmented speech.
	\item Define a speech augmentation based unsupervised learning approach, utilizing the similarity between the bottleneck layer feature, as well as the audio reconstructing information for auxiliary training.
\end{itemize}

The rest of the paper is organized as follows. Sec.~\ref{sec:related} highlights
the related prior works about data augmentation, unsupervised learning, and other methodologies of KWS tasks. Sec.~\ref{sec:method} describes the proposed model architecture and augmentation based unsupervised learning loss. Sec.~\ref{sec:exp} reports the experimental results compared with other pre-training methods. We also discuss relationship between pre-training steps and performance of downstream KWS tasks. In Sec.~\ref{sec:conclu}, we conclude with the summary of the paper and future works.

\section{Related Work}
\label{sec:related}
Data augmentation is a common strategy to enlarge the training set of speech applications, such as Automatic Speech Recognition (ASR) and Keyword Spotting (KWS). The work~\cite{jaitly2013vocal} studied the vocal tract length perturbation method to improve the performance of ASR systems. The work~\cite{ko2015audio} investigated a speed-perturbation technique to change the speed of the audio signal. Noisy audio signals have been used in \cite{hannun2014deep}, corrupting clean training speech with noise signal, to improve the robustness of the speech recognizer. SpecAugment~\cite{park2019specaugment} is a spectral-domain augmentation whose effect is to mask bands of frequency and/or time axes. SpecAugment is also explored further on large scale dataset in \cite{park2020specaugment}. WavAugment~\cite{kharitonov2021data} combines pitch modification, additive noise and reverberation to increase the performance of Contrastive Predictive Coding (CPC). In this work, we apply the speed and volume perturbation in our speech augmentation method.

Although supervised learning has been the major approach in keyword spotting area, current supervised learning models require large amounts of labeled data. 
Those high quality labeled datasets require substantial effort and are hardly available for the less frequently used languages. For this reason, recently there has been a great surge of interest in weakly supervised solutions that use datasets with few human annotations. Noisy student training, a semi-supervised learning method was proposed to ASR~\cite{park2020improved} and later used for robust keyword spotting~\cite{park2021noisy}. There also have been related researches investigating the use of unsupervised methods to perform keyword spotting~\cite{garcia2006keyword,li2007novel,zhang2009unsupervised}. \cite{garcia2006keyword} proposed a self-organizing speech recognizer, and minimal transcriptions are used to train a grapheme-to-sound-unit converter. \cite{li2007novel} presented a prototype KWS system that doesn't need manually transcribed data to train the acoustic model. In \cite{zhang2009unsupervised}, the authors proposed an unsupervised learning framework without transcription. A GMM model is used to label keyword samples and test utterances by Gaussian posteriorgram. After that, segmental dynamic time warping (SDTW) gives a relevant score, and ranks the score to figure out the output. The feasibility and effectiveness of these results encourage us to introduce unsupervised learning framework to the task of keyword spotting.

Google Speech Commands V2 Dataset, is a well-studied and benchmarked dataset for novel ideas in KWS. A lot of previous works perform experiments on this dataset. \cite{de2018neural} introduced a convolutional recurrent network with attention on multiple KWS tasks. MatchboxNet~\cite{majumdar2020matchboxnet} is a deep residual network composed from 1D time-channel separable convolution, batch-norm layers, ReLU and dropout layers. Inspired by \cite{de2018neural} and \cite{majumdar2020matchboxnet}, EdgeCRNN~\cite{wei2021edgecrnn} was proposed, an edge-computing oriented model of acoustic feature enhancement for keyword spotting. Recently, \cite{vygon2021learning} combined a triplet loss-based embedding and a variant of K-Nearest Neighbor (KNN) for classification. 
We also evaluated our speech augmentation based unsupervised learning method on this dataset, and compared with other unsupervised approaches, including CPC~\cite{oord2018representation}, APC~\cite{chung2019unsupervised} and MPC~\cite{Improving}.

\section{Proposed Method}
\label{sec:method}

\subsection{KWS Model Architecture}
\begin{figure}[htbp]
\centerline{\includegraphics[width=0.82\columnwidth]{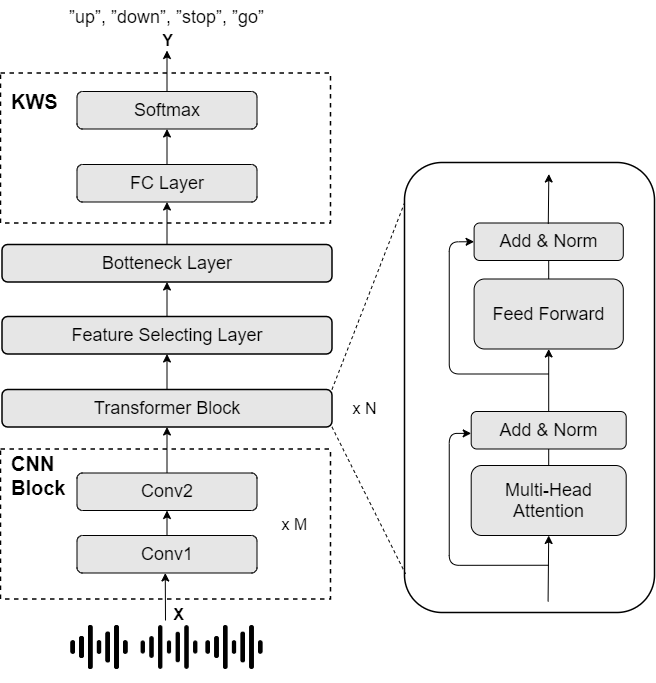}}
\caption{The Architecture of our CNN-Attention model for keyword spotting task. The network is composed of CNN layers, self-attention layers, feature selecting layer, bottleneck layer, and project layer. In the feature selecting layer, the last few frames are selected. Finally, the project layer maps the features to predict the keyword classification.}
\label{fig}
\end{figure}

The keyword spotting task could be described as a sequence classification task. The keyword spotting network maps an input audio sequence $X=(x_0, x_1, ..., x_T)$ to a limited of keyword classes $Y\in y_{1:S}$. In which, $T$ is the number of audio frames and $S$ is the number of classes. Our proposed model architecture for keyword spotting is shown in Fig~\ref{fig}. The network contains five parts: (1) CNN Block, (2) Transformer Block, (3) Feature Selecting Layer, (4) Bottleneck Layer, and (5) Project Layer.

The CNN block consists of several 2D-convolutional layers, handling the local variance on time and spectrum axes.\begin{equation} 
E_{cnn} = \mathrm{2DConv}_{\times N}(X)
\end{equation}
In which, $N$ is the number of convolutional layers. Then, the CNN output $E_{cnn}$ is inputted to the transformer block, to capture long-time information with self-attention mechanism.
\begin{equation} 
E_{tran} = \mathrm{SelfAttention}_{\times M}(E_{cnn})
\end{equation}
In which, $M$ is the number of self-attention layers. After transformer block, we designed a feature selecting layer to extract keyword information from sequence $E_{tran}$.
\begin{equation} 
E_{feat} = \mathrm{Concat}(E_{tran}[T-r,T])
\end{equation}
In feature selecting layer, we firstly collect last $r$ frames of $E_{tran}$. And then, we concatenate all the collected frames together, into one feature vector $E_{feat}$. After feature selecting layer, we use a bottleneck layer and a project layer, projecting the hidden states to the predicted classification classes $\tilde{Y}$.
\begin{equation} 
E_{bn} = \mathrm{FC_{bn}}(E_{feat})
\end{equation}
\begin{equation} 
\tilde{Y} = \mathrm{FC_{proj}}(E_{bn})
\end{equation}
Finally, the the cross-entropy (CE) loss for supervised learning and model fine-tuning is calculated via predicted classes $\tilde{Y}$ and ground truth classes $Y$.  
\begin{equation} 
\mathcal{L}_{ce} = \mathrm{CE}(Y, \tilde{Y})
\end{equation}

\subsection{Augmentation Method}
Data augmentation are the most common used methods to promote the robustness and performance of the model in speech tasks. In this work, speed and volume based augmentation are investigated in the unsupervised learning of keyword spotting. For a given audio sequence $X$, we denote it as the amplitude $A$ and time index $t$.
\begin{equation} 
X = A(t)
\end{equation}
For speed augmentation, we set a speed ratio $\lambda_{speed}$ to adjust the speed of $X$.
\begin{equation} 
X^{aug} = A(\lambda_{speed}t)
\end{equation}
For volume augmentation, we also set an intensity ratio $\lambda_{volume}$ to change the volume of $X$.
\begin{equation} 
X^{aug} = \lambda_{volume}A(t)
\end{equation}
With different ratios $\lambda_{speed}$ and $\lambda_{volume}$, we could obtain multiple speech sequence pairs $(X, X^{aug})$, to train the audio representation network with unsupervised learning. We think that speech utterances at different speed or volume should have similar high-level feature representation for KWS tasks.


\begin{table*}[b]
	\centering 
	\caption{Model Configurations}
	\label{tab1}
	\begin{tabular}{p{5cm}|p{10cm}<{\centering}}
		\hline
		\hline
		\textbf{Unit Name} & \textbf{Hyperparameters}\\ 
		\hline 
		\#CNN Blocks & $M = 2$ layers, $3\times 3$ kernel, $2\times 2$ stride, $32$ channels \\  
		\hline
		\#Transformer Block & $N = 2$ layers, dimension = $320$, $4$ head, feedforward = $1024$ \\      
		\hline
		\#Feature Selecting Layer & Last $r = 2$ frames, $2 \times 320$ dimension \\      
		\hline
		\#Bottleneck Layer & one FC layer, $800$ dimension \\      
		\hline 
		\#Project Layer & one FC layer, $12$ dimension softmax\\      
		\hline
		\#Reconstruct Layer & one FC layer, $40$ dimension softmax\\
		\hline
		\#Factor Ratio & $\lambda_1=0.9$, $\lambda_2=0.05$, $\lambda_3=0.05$ \\
		\hline
		\hline
	\end{tabular}  
\end{table*}

\subsection{Unsupervised Learning Loss}

\begin{figure}[ht]
	\centerline{\includegraphics[width=0.82\columnwidth]{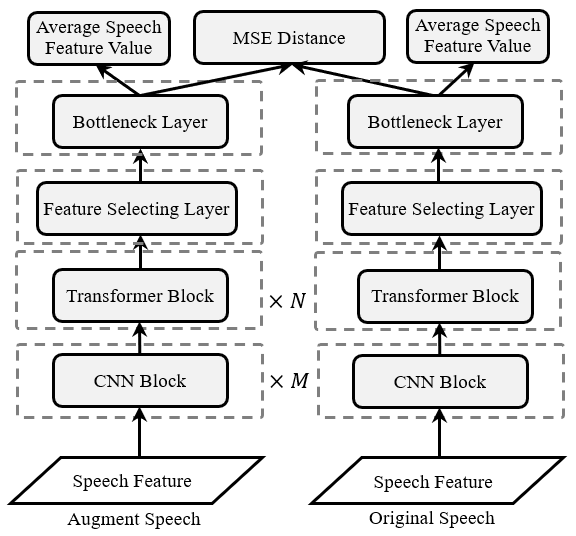}}
	\caption{The proposed speech augmentation based audio unsupervised learning method. In the pre-training stage, the pair of original and augmented speech will be inputted into the network separately but with the same model parameters. The network will output the average speech feature values and the bottleneck feature. The two bottleneck features are calculated by MSE loss, since the augmented and original speech should output similar high-level features for keyword spotting.}
	\label{fig2}
\end{figure}

The overall architecture of augmentation based unsupervised learning is shown in Fig~\ref{fig2}. Similar to other unsupervised methods, the proposed approach also consists of two stages: (1) pre-training on unsupervised data, and (2) fine-tuning on supervised KWS data. In the pre-training stage, the bottleneck feature was obtained through training the unlabeled speech. In fine-tuning stage, the extracted bottleneck features are used for KWS prediction.

In the pre-training stage, the pair speech data $(X, X^{aug})$ are inputted into the CNN-Attention models respectively, but with the same model parameters. Because $X^{aug}$ comes from $X$, our designed unsupervised methods expect that $X$ and $X^{aug}$ will output similar high-level bottleneck features. It means that no matter how fast or how loud a speaker says, the content of the speech is the same. Thus, the optimization of network needs to reflect the similarity of $X$ and $X^{aug}$. We choose the Mean Square Error (MSE) $\mathcal{L}_{sim}$ to measure the distance between the output of $X$ and $X^{aug}$.
\begin{equation} 
\begin{aligned}
\mathcal{L}_{sim} = \frac{1}{U_{bn}}\sum_{u=0}^{U_{bn}}|E_{bn}(u) - E^{aug}_{bn}(u)|^2
\end{aligned}
\end{equation}
Where $U_{bn}$ represents the dimension of the bottleneck feature vector. $E_{bn}$ and $E^{aug}_{bn}$ are the output of bottleneck layer of original speech $X$ and augmented speech $X^{aug}$ respectively.

In addition, the designed network has another branch for auxiliary training, which predicts the average feature of the input speech segment. This branch guides the network to learn the intrinsic feature of the speech utterance. We firstly compute the average vector of the input Fbank vector $X$ alongside the time axis $t$. Then, we use another reconstructing layer attached to the bottleneck layer, to reconstruct the average Fbank vector $\tilde{X}$. We also use MSE loss $\mathcal{L}_{x}$ to calculate the similarity between these two audio vectors alongside the feature dimension $U_{x}$.
\begin{equation} 
\begin{aligned}
\mathcal{X} &= \frac{1}{T}\sum_{T}(X) \\
\tilde{X} &= \mathrm{FC_{reconstruct}}(E_{bn}) \\
\mathcal{L}_{x} &= \frac{1}{U_{x}}\sum_{u=0}^{U_{x}}|\mathcal{X}(u) - \tilde{X}(u)|^2
\end{aligned}
\end{equation}

In which, $U_{x}$ represents the dimension of Fbank feature vector, and $\mathcal{X}$ denotes the average vector of $X$. Similarly, the loss $\mathcal{L}^{aug}_{x}$ between the augmented average audio $\mathcal{X}^{aug}$ and ured feature $\tilde{X}^{aug}$ could be defined as follows:
\begin{equation} 
\begin{aligned}
\mathcal{L}^{aug}_{x} &= \frac{1}{U_{x}}\sum_{u=0}^{U_{x}}|\mathcal{X}^{aug}(u) - \tilde{X}^{aug}(u)|^2
\end{aligned}
\end{equation}

Therefore, the final loss function $\mathcal{L}_{ul}$ of the unsupervised learning (UL) consists of the above three losses $\mathcal{L}_{sim}$, $\mathcal{L}_{x}$, and $\mathcal{L}^{aug}_{x}$.
\begin{equation}
	\begin{aligned}
		\mathcal{L}_{ul} = \lambda_1\mathcal{L}_{sim} + \lambda_2\mathcal{L}_{x} + \lambda_3\mathcal{L}^{aug}_{x}
	\end{aligned}
\end{equation}
Where $\lambda_1$, $\lambda_2$, $\lambda_3$ are factor ratio of each loss component.

In fine-tuning stage, the branch of average feature prediction is removed. A project layer and a softmax layer are added after the bottleneck layer to make the KWS prediction. In the fine-tuning, the parameters of original network could be fixed or updated. In our experiments, we found that updating all the parameters could help to improve the performance. Thus, we choose to update all parameters in the fine-tuning stage.

\section{Experiments}
\label{sec:exp}

In this section, we evaluated the proposed method in keyword spotting tasks. We implemented our CNN-Attention model with supervised training and compared it with Google's model. We also made an ablation study, to explore the effect of speed and volume augmentation on unsupervised learning. What's more, other unsupervised learning methods are compared with our approach, including CPC, APC, MPC. When implementing these approaches, we used the network and hyperparameters in their publications, but all experimental tricks were not leveraged\cite{Improving, chung2019unsupervised, oord2018representation}. We also discuss the impact of different pre-training steps on the performance and convergence of downstream KWS task.

\subsection{Datasets}
We used Google’s Speech Commands V2 Dataset~\cite{warden2018speech} for evaluating the proposed models. The dataset contains about $106000$ one-second or more long utterances. Total $30$ short words were recorded by thousands of different people, as well as background noise such as pink noise, white noise, and human-made sounds. The KWS task is to discriminate among $12$ classes: ``yes'', ``no'', ``up'', ``down'', ``left'', ``right'', ``on'', ``off'', ``stop'', ``go'', unknown, or silence. The dataset was split into training, validation, and test sets, with $80\%$ training, $10\%$ validation, and $10\%$ test. This results in about $37000$ samples for training, and $4600$ each for validation and testing. We used the real noisy data HuNonspeech\footnote{http://web.cse.ohio-state.edu/pnl/corpus/HuNonspeech/} to corrupt the original speech. In the experiments, the Aurora4 tools were used to implement this strategy\footnote{http://aurora.hsnr.de/index-2.html}. Each utterance will be randomly corrupted by public $100$ kinds of noise in HuNonspeech. Each utterance has a level of $0$-$20$dB Signal Noise Ratio (SNR), and all datasets have an average $10$dB SNR. 

Similar to other unsupervised methods, a large unlabeled corpus, $100$ hours of Librispeech~\cite{Panayotov2015Librispeech} clean speech were also leveraged to pre-train the network by unsupervised learning. Firstly, the long utterances were split up into $1$ second segments, keeping consistent with Speech Commands datasets. Nextly, the clean segments were also mixed with noisy HuNonspeech data by Aurora 4 tools, and the corrupted mechanism was as same as the Speech Commands. 

\subsection{Experimental Setups}
\begin{table*} [ht] 
	\centering 
	\caption{Results Comparison of KWS Model, Classification Accuracy (\%)}
	\label{tab2}
	\begin{tabular}{p{8cm}p{4cm}p{1cm}p{1cm}}  
		\hline
		\hline
		Model Name & Supervised Training Data & Dev & Eval\\ 
		\hline
		Sainath and Parada (Google) & Speech Commands & - & 84.7 \\   
		CNN-Attention (ours) & Speech Commands & 86.4 & 85.3\\    
		\textbf{CNN-Attention + volume \& speed augment (ours)} & Speech Commands & \textbf{87} & \textbf{85.7} \\
		\hline 
		\hline 
	\end{tabular}  
\end{table*}

\begin{table*} 
	\centering 
	\caption{Ablation Study, the effect of speed and volume augmentation, Classification Accuracy (\%)}
	\label{tab3}
	\begin{tabular}{p{7cm}p{3cm}p{3cm}p{1.5cm}p{1.5cm}}  
		\hline
		\hline
		Model Name & Pre-training Data & Fine-tuning Data & Dev & Eval\\  
		\hline
		CNN-Attention + volume pre-training & Speech Commands & Speech Commands & 86.1 & 85.9\\     
		CNN-Attention + speed pre-training & Speech Commands & Speech Commands & 87.8 & 86.9 \\ 
		\textbf{CNN-Attention + volume \& speed pre-training} & Speech Commands & Speech Commands & \textbf{87.9} & \textbf{87.2} \\
		\hline
		CNN-Attention + volume pre-training & Librispeech-100 & Speech Commands & 86.3 & 86.0\\     
		CNN-Attention + speed pre-training & Librispeech-100 & Speech Commands & 87.9 & 87.9 \\ 
		\textbf{CNN-Attention + volume \& speed pre-training} & Librispeech-100 & Speech Commands & \textbf{88.2} & \textbf{88.1} \\
		\hline 
		\hline
	\end{tabular}  
\end{table*}

The acoustic features were $40$-dimensional log-mel filterbank with $30$ms frame length and $10$ms frame shift. The detailed hyperparameters of our proposed network were shown in Table~\ref{tab1}. For training the KWS model, all of the matrix weights are initialized with random uniform initialization, and the bias parameters are initialized with the constant value $0.1$. In our experiments, we trained all the networks with Adam optimizer for $30$k steps with a batchsize $200$ until the loss becomes little change. In addition, the factor ratios of loss $\lambda_1$, $\lambda_2$, and $\lambda_3$ are set to $0.9$, $0.05$, $0.05$ respectively.

To demonstrate the effectiveness of our proposed model, we investigated several other approaches for comparison. For supervised learning, we used Sainath and Parada’s model by Google~\cite{sainath2015convolutional} as the baseline model. The Google blog post released the Sainath and Parada’s model implemented by TensorFlow. For unsupervised learning, we compared our method with other pre-training models:

\begin{itemize}
	\setlength{\itemsep}{0pt}
	\setlength{\parsep}{0pt}
	\setlength{\parskip}{0pt}
	\item \textbf{Contrastive Predictive Coding (CPC)~\cite{oord2018representation}}: Through an unsupervised mechanism by utilizing next step prediction, CPC learns representations from high-dimensional signal. The CPC network mainly contains a non-linear encoder and an autoregressive decoder. An input sequence is embedded to a latent space, producing a context representation. Targeting at predicting future observations, the density ratio is established to maximize the mutual information between future observations and current context representation.
	\item \textbf{Autoregressive Predictive Coding (APC)~\cite{chung2019unsupervised}}: APC also belongs to the family of predictive models. APC directly optimizes L1 loss between input sequence and output sequence. APC has proved an effective method in recent language model pre-training task and speech representation.
	\item \textbf{Masked Predictive Coding (MPC)~\cite{Improving}}:
	Inspired by BERT, MPC uses Masked Language Model (MLM) structure to perform predictive coding on Transformer based models. Similar to BERT, $15\%$ of feature frames in each utterance are chosen to be masked during the pre-training procedure. Among these chosen frames, $80\%$ are replaced with zero vectors, $10\%$ are replaced with random positions, and the rest remain unchanged. L1 loss is computed between masked input features and encoder output at corresponding position. Dynamic masking was also adopted where the masking pattern is generated when a sequence is fed into the model.
\end{itemize}



\subsection{Results}

\begin{table*} 
	\centering 
	\caption{Compared with Other Unsupervised Learning Methods, Classification Accuracy (\%)}
	\label{tab4}
	\begin{tabular}{p{8cm}p{3cm}p{3cm}p{1.0cm}p{1.0cm}}  
		\hline
		\hline
		Model Name & Pre-training Data & Fine-tuning Data & Dev & Eval\\  
		\hline
		Contrastive Predictive Coding (CPC)~\cite{oord2018representation} & Speech Commands & Speech Commands & 87.6 & 86.9 \\     
		Autoregressive Predictive Coding (APC)~\cite{chung2019unsupervised} & Speech Commands & Speech Commands & 87.2 & 86.5 \\
		Masked Predictive Coding (MPC)~\cite{Improving} & Speech Commands & Speech Commands & 87.0 & 86.7 \\
		\textbf{CNN-Attention + volume \& speed pre-training (ours)} & Speech Commands & Speech Commands & \textbf{87.9} & \textbf{87.2} \\
		\hline
		Contrastive Predictive Coding (CPC)~\cite{oord2018representation} & Librispeech-100 & Speech Commands & 87.8 & 87.4 \\     
		Autoregressive Predictive Coding (APC)~\cite{chung2019unsupervised} & Librispeech-100 & Speech Commands & 87.7 & 87.5 \\
		Masked Predictive Coding (MPC)~\cite{Improving} & Librispeech-100 & Speech Commands & 87.9 & 87.0 \\
		\textbf{CNN-Attention + volume \& speed pre-training (ours)} & Librispeech-100 & Speech Commands & \textbf{88.2} & \textbf{88.1} \\
		\hline 
		\hline
	\end{tabular}  
\end{table*}

Table~\ref{tab2} lists the experimental results of supervised learning with Speech Commands dataset. We firstly implemented the Google's Sainath and Parada model by the original TensorFlow recipes, achieving the accuracy of $84.7\%$. Secondly, our CNN-Attention model is implemented by supervised loss $\mathcal{L}_{ce}$ without any augmented data and achieved $0.6\%$ higher accuracy than Google's model. It is proved that our designed CNN-Attention architecture is effective for KWS task. Finally, after adding speed and volume augmentation to speech, we got a higher accuracy. It corresponds with the existing research that augmented dataset is helpful for improving the performance of the model. It also inspires our motivation for building augmentation based unsupervised learning methods.

To analyze the effect of speed and volume augmentation on unsupervised learning, we also made an ablation study in our experiments. The experimental results are shown in Table~\ref{tab3}. The volume pre-training model means that the augmented speech pairs $(X, X^{aug})$ only contain the intensity augment data. Meanwhile, the speed pre-training model is trained only by speed augmented pairs. For better investigation, we pre-trained the model with two datasets by unsupervised learning loss $\mathcal{L}_{ul}$. The results indicate that speed augmented unsupervised learning has better performance than intensity based augmented pre-training. With both volume and speed augmentation, we could achieve better classification accuracy than only with single augmentation method. In addition, large datasets pre-training (Librispeech-100) results in better performance than small datasets (Speech Commands). Our proposed augmentation based unsupervised method (Eval $87.2\%$ in Table~\ref{tab3}) also promotes the accuracy of adding augmentation to supervised training (Eval $85.7\%$ in Table~\ref{tab2}) even with the same training data.

After that, we established the CPC, APC, MPC and made the comparison with these unsupervised learning methods. As depicted in Table~\ref{tab4}, CPC achieves better performance than APC and MPC. Our augmentation based approach outperforms all of the other unsupervised methods on both two pre-training datasets (Speech Commands and Librispeech-100). The comparison demonstrated that our proposed augmentation based unsupervised learning is capable of extracting the speech information, and is an effective approach for KWS tasks. 

\subsection{Pre-training Analysis}

\begin{figure}[ht] 
	\centerline{\includegraphics[width=1.0\columnwidth]{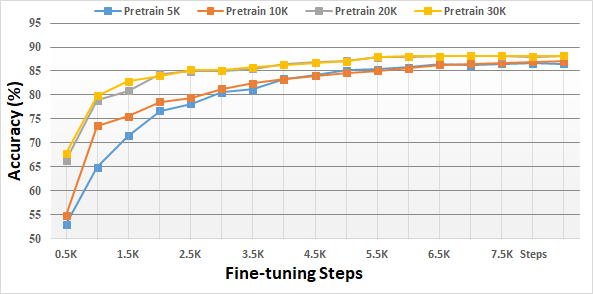}}
	\caption{The results comparison with different pre-training steps. Different pre-training steps of unsupervised learning result in different accuracy performance and fine-tuning convergence. In our experiments, pre-training $30K$ steps have the highest classification accuracy, and fastest convergence.}
	\label{fig3}
\end{figure}

More pre-training steps usually help to improve the performance of downstream tasks. To get a better understanding of our unsupervised approach, we also conducted experiments with different pre-training steps. The $5K, 10K, 20K, 30K$ pre-training steps were used for making this comparison. The performance of different steps is plotted in Fig~\ref{fig3}.

We show the model training of supervised learning with these different steps of pre-training. Our experiments demonstrated that more pre-training steps are not only helpful for achieving better performance but also making downstream KWS task converge faster. Unsupervised learning with $30K$ steps has the highest classification accuracy and the fastest convergence. It also should be noted that the difference between $20K$ and $30K$ was very close, meaning that the pre-training steps are enough to obtain the desired performance.



\section{Conclusion}
\label{sec:conclu}

This paper investigated unsupervised learning method for keyword spotting task. We designed a CNN-Attention architecture and achieved competitive results on the Speech Commands dataset. In addition, we proposed a speech augmentation based unsupervised learning approach for KWS. Our method uses speed and intensity augmentation to establish training pairs, and pre-trains the network via the similarity loss between the speech pair and the speech reconstructed loss. In our experiments, the proposed unsupervised approach could further improve the model performance, and outperform other unsupervised methods, such as CPC, APC and MPC. We also found that more pre-training steps are not only helpful for better performance but also for faster convergence. In future works, we are interested in applying the augmentation based unsupervised learning approach to other speech tasks, such as speaker verification and speech recognition.

\section{Acknowledgement}
\label{sec:ack}
This paper is supported by the Key Research and Development Program of Guangdong Province under grant No.
2021B0101400003. Corresponding author is Jianzong Wang from Ping An Technology (Shenzhen) Co., Ltd (jzwang@188.com).

\vfill\pagebreak
\bibliographystyle{IEEEtran}
\bibliography{IJCNN2021_KWS}

\end{document}